\documentclass[reprint,12pt,authoryear]{elsarticle}

\journal{Radiation Physics and Chemistry}

\usepackage{amsmath,amscd,amssymb}
\usepackage{bm}

\usepackage{graphicx}
\usepackage{lineno,hyperref}

\begin{document}
\begin{frontmatter}

\title{Energy loss reduction of a charge moving through an anisotropic plasma-like medium}

\author[]{Aleksandra A.~Grigoreva\corref{mycorrespondingauthor}}
\cortext[mycorrespondingauthor]{Corresponding author}
\ead{a.a.grigoreva@spbu.ru}

\author{Andrey V. Tyukhtin}
\author{Sergey N.~Galyamin}
\author{Tatiana Yu.~Alekhina}

\address{Saint Petersburg State University, 7/9 Universitetskaya nab., St. Petersburg, 199034 Russia}

\begin{abstract}
We analyze radiation of a charge moving in a vacuum channel in an anisotropic non-gyrotropic medium with plasma-like components of the permittivity tensor. The expressions for field components are obtained and analyzed. It is shown that the field contains both the radiation field and the plasma oscillations. Most attention is focused on the energy loss of the charge per the unit path length. The dependencies of the loss on the charge velocity and the plasma frequencies are studied. The relative roles of radiation loss and polarization one are considered. The most interesting result is that the energy loss is negligible when one of the permittivity tensor components is equal to 1, and the charge velocity tends to the speed of light in vacuum. This effect can be promising for applying in collimators of ultrarelativistic bunches.
\end{abstract}

    
\end{frontmatter}
    

\section{Introduction}

Interaction of a charged particle bunch with different slow-wave structures (including uniform dielectric media) results in generation of various types of radiation. This radiation takes the energy away of the bunch, i.~e. causes the loss of the bunch energy. In many devices, such as radiation sources (classical vacuum microwave and THz devices, as well as modern FELs), detectors and bunch diagnostis systems, this effect is considered as positive. However, there is another class of devices where the mentioned issue is parasitic and should be minimized. This is especially the case for modern particle accelerators where the interaction of the bunch with various beamline structures results in undesired decrease in the bunch quality including the particle energy.

Typically, the large portion of parasitic effects in accelerators and colliders is connected with its collimation system~\citep{Novokhatski14, AntipovSA20}.
Therefore, new materials are considered for traditional collimator assembly~\citep{AntipovSA20} and alternative dielectric-based collimation systems are also discussed~\citep{Kanareykin10, SchoessowKanareykin12}. Moreover, the successful usage of hollow electron lenses (low-energy hollow electron beams) for efficient halo removal of intense high-energy beams in storage rings and colliders should be particularly mentioned~\citep{Stancari11, Fischer15, Gu20}. It is equally important that similar hollow electron structures (called in that context the hollow plasma channels) were utilized recently in successful experiments on plasma wakefield acceleration \citep{Blum07, Gessner16}.

When a charged particle bunch passes through a channel in a medium, Cherenkov radiation (CR) is generated and causes the corresponding radiation loss. The CR theory for the case of an infinite medium is well known~\citep{B62}. It should be underlined that an increase in charge velocity, as a rule, increases the radiation loss per unit path length of the charge. This effect takes place for both nondispersive isotropic medium and isotropic medium having typical frequency dispersion; however, it is interesting that this increase is slower in the second case~\citep{Tyukhtin05}.

The main aim of the present paper is to show that the specific type of anisotropy and dispersion of the medium results in a strong decrease of the CR loss for ultrarelativistic bunches. This fact allows to minimize the energy loss at collimation of the ultrarelativistic bunches.

We consider the electromagnetic field of a charge moving through a vacuum channel in an anisotropic non-gyrotropic uniaxial medium. It is assumed that components of a permittivity tensor possess plasma-like frequency dispersion.  
As it will be shown these components have different signs in the frequency range which is significant for radiation, i.~e. the medium is so-called ``hyperbolic medium'' within this range.

Such a medium can be implemented in different ways. One of them is the use of metamaterials~\citep{SZKKE08} which are considered very prospective during last decades due to a wide range of interesting possibilities they provide for physics and techniques: negative refraction, focusing effect etc.~\citep{NaturePhoton, HMM_PQE, SciRep, Thzfocusing}.
For the goals of this study, the so-called ``hyperbolic metamaterials'' (HMM) are appropriate, for example, a HMM based on a two-dimensional silicon pillar array microstructure~\citep{Thzfocusing}. Another suitable example is a ``wire medium'', an artificial lattice comprized of long metal conductors with small spacings~\citep{BTV02, TD11}. Unfortunately, the latter structure possesses parasitic spatial dispersion which, however, can be suppressed~\citep{DP08}. In recent years, metamaterials has attracted an essential interest in the context of their implementation in particle beam physics, including wakefield generation for particle detector design~\citep{Ant08}
and modern high-gradient accelerators~\citep{HAnd18, LuShapConde19}.

A more traditional way to implement the medium with the desired properties is the use of an electron plasma or an electron flux placed in a strong constant magnetic field. If the electron gyrofrequency is much larger than the plasma frequency and the typical frequencies of generated radiation then the permittivity tensor is close to the desired diagonal tensor. The successes noted above in the use of the hollow electron beams~\citep{Stancari11, Fischer15, Gu20, Blum07, Gessner16} allow us to hope for the successful use of the method proposed here to reduce the energy loss of bunches subjected to collimation.  

\section{Electromagnetic field of a charge}

\begin{figure}[h]
\centering
\includegraphics[scale=1]{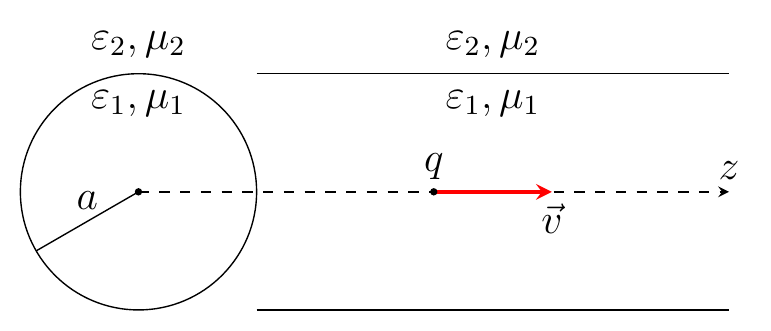}
\caption{Geometry of the problem: the transverse (left) and longitudinal (right) cross-sections.}
\label{geometry}
\end{figure}

The system under study is an anisotropic uniaxial medium possessing a vacuum channel with radius $a$. The anisotropy axis coincides with the channel axis ($z$-axis, see Fig.~\ref{geometry}). The permittivities ($\varepsilon$) and permeabilities  ($\mu$) in the channel and in the external area are
\begin{equation}
\varepsilon_1= \mu_1=\mu_2=1, \ \ \
    \hat{\varepsilon}_2 =
    \begin{pmatrix}
        \varepsilon_\perp & 0 & 0 \\
        0 & \varepsilon_\perp & 0 \\
        0 & 0 & \varepsilon_\parallel
    \end{pmatrix}.
\label{e-tensor}
\end{equation}
Point charge moves along the channel axis with constant velocity $\vec{v}=v\vec{e}_z$. The charge and current densities are
\begin{equation}
    \label{charge_density}
    \rho = q \delta\left(x\right) \delta\left(y\right) \delta\left(z-vt\right), \; \vec{j} = v \rho \vec{e}_z.
\end{equation}
Note that electromagnetic field structure in the case of the charge intersecting a boundary between a vacuum and an anisotropic medium with characteristics (\ref{e-tensor}) was analyzed in \citep{GT11}, however the problem of energy loss was not considered and the channel radius was not taken into account.

The general form of the electromagnetic field components can be obtained using the Fourier transform~\citep{B62}. Here we give only the final expressions for the longitudinal component of the electric field:
\begin{equation}
\label{field_integral}
E_z  = \int\nolimits_{-\infty}^{+\infty} e_{z\omega} \left( \omega \right)
\exp \left( i \zeta \frac{\omega}{v} \right) d\omega,
\end{equation}
where
\begin{equation}
e_{z \omega} \left( \omega \right) = \frac{i q \kappa^2}{\pi \omega}
\left\{
        \begin{aligned}
            & - K_0\left( \kappa r \right) + \frac{F_1 \left(\omega\right) }{F_2 \left(\omega\right) } I_0\left( \kappa r \right) \text{ for } r \leq a, \\
            & H_0^{\left( 1 \right)} \left( s r \right) \frac{s}{\kappa a F_2 \left(\omega\right)} \text{ for } r > a,
        \end{aligned}
\right.
\label{ez_fourier}
\end{equation}
\begin{equation}
\begin{aligned}
&F_1 \left(\omega\right) = \varepsilon_\parallel \kappa K_0\left( \kappa a \right) H_1^{\left( 1 \right)} \left( sa \right) +
s H_0^{\left( 1 \right)} \left( sa \right) K_1\left( \kappa a \right),   \\
&F_2 \left(\omega\right)  = \varepsilon_\parallel \kappa I_0\left( \kappa a \right) H_1^{\left( 1 \right)} \left( sa \right) - s H_0^{\left( 1 \right)} \left( sa \right) I_1\left( \kappa a \right) , 
\end{aligned}
\label{F1F2}
\end{equation}
\begin{equation}
\kappa^2 = \frac{ \omega^2}{v^2}  \left( 1 - \beta^2 \right) , \; s^2 = \frac{\omega^2}{v^2} \frac{\varepsilon_\parallel}{\varepsilon_\perp} \left( \varepsilon_\perp \beta^2 -1 \right), \; \zeta = z - vt, \;  \beta = v/c.
\label{ks}
\end{equation}
Here $c$ is the light velocity in a vacuum; $ i \kappa \left(\omega\right)$ and $s\left(\omega\right)$ are orthogonal components of the wave vector in the channel and in the exterior area, respectively. The rule for the square root extraction
$\left( s=\sqrt{s^2\left(\omega\right)} \right)$ is the following: the branch cuts coincide with lines $\mbox{Im} \left( s\left(\omega\right) \right) = 0$ and the physical Riemann surface sheet is fixed by the requirement $\mbox{Im}\left(s\left(\omega\right)\right) \geq 0$ for $\omega \in \mathbb{R}$.

We consider the case when the permittivity tensor components have the plasma-like form:
\begin{equation}
    \varepsilon_\parallel = 1 - \frac{\omega_{p\parallel}^2}{\omega^2 + i \nu_\parallel \omega} , \ \ \
    \varepsilon_\perp = 1 - \frac{\omega_{p\perp}^2}{\omega^2 + i \nu_\perp \omega},
\label{epsilons}
\end{equation}
where $\omega_{p\parallel}$, $\omega_{p\perp}$ are plasma frequencies and $\nu_\parallel$, $\nu_\perp$ are values responsible for losses. Further we will consider the loss-free medium (i.~e. $\nu_{\parallel,\perp} \to 0$), infinitesimal losses will be used to establish the location of singularities on the complex plane $\omega$. We restrict the analysis to the case $\omega_{p\perp} < \omega_{p\parallel}$ because the most interesting situation is realized when $\omega_{p\perp} \to 0$.

\begin{figure*}
    \begin{minipage}{0.49\linewidth}
        \centering
        \includegraphics[width=\linewidth]{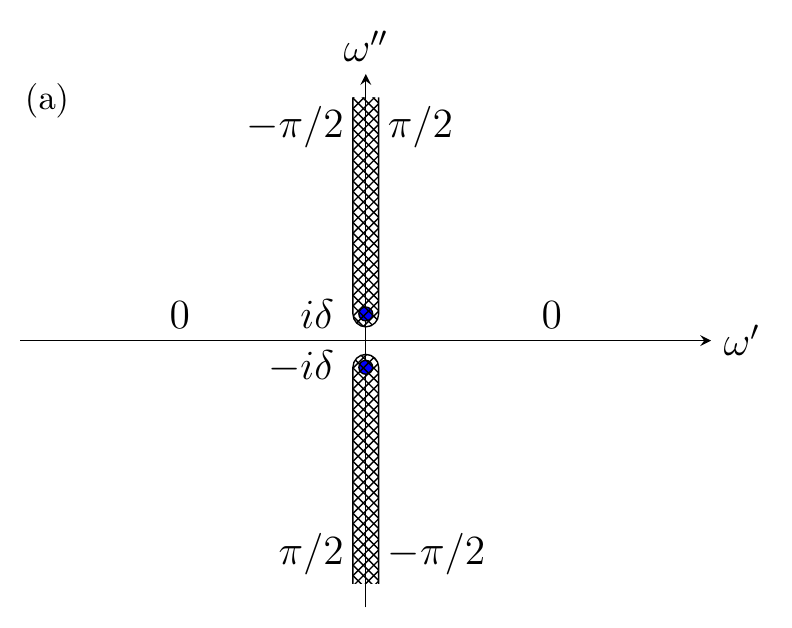}
    \end{minipage}
    \hfill
    \begin{minipage}{0.49\linewidth}
        \centering
        \includegraphics[width=\linewidth]{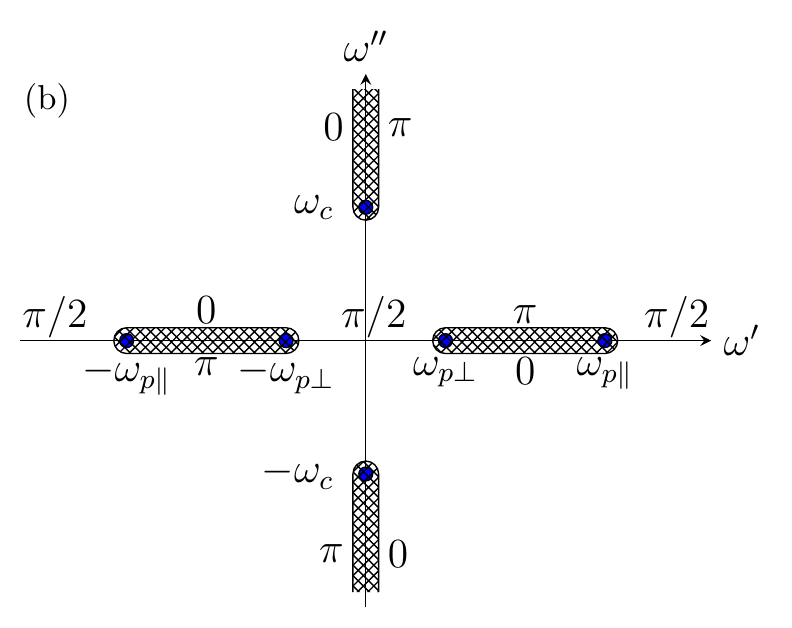}
    \end{minipage}
    \caption{Location of branch points (blue circles) and branch cuts (crosshatched strips) of functions $\kappa \left(\omega\right)$ (a), $s\left(\omega\right)$ (b) on the complex plane $\omega$. Numbers mean the function arguments on the real axis and on the branch cuts.}
    \label{transfunctions_complex}
\end{figure*}

By the above-mentioned rule, the functions $\kappa \left( \omega \right)$ and $s\left(\omega\right)$ take the following form:
\begin{equation}
\label{trans_functions}
\kappa \left( \omega \right) = \left| \kappa \left( \omega \right) \right| \exp{ \left( i \arg \left( k \left( \omega \right) \right) \right) }, \;
s \left(\omega\right) = \left| s \left( \omega \right) \right| \exp{ \left( i \arg\left( s \left(\omega\right) \right) \right) },
\end{equation}
\begin{equation}
\begin{aligned}
\left| \kappa \left( \omega \right) \right| &= \frac{ \sqrt{1 - \beta^2} }{v} \sqrt{ \left| \omega - i \delta \right| \left| \omega + i \delta \right| }, \\
\left| s \left( \omega \right) \right| &= \frac{\sqrt{ 1 - \beta^2 }}{v} \sqrt{ \frac{ \left| \omega - \omega_{p\parallel} \right| \left| \omega + \omega_{p\parallel} \right| }{ \left| \omega - \omega_{p\perp} \right| \left| \omega + \omega_{p\perp} \right| } } \sqrt{ \left| \omega - \omega_c \right| \left| \omega + \omega_c \right| },
\end{aligned}
\end{equation}
where $\delta$ is real positive infinitesimal value, $\omega_c = i \omega_{p\perp} \beta / \sqrt{ 1 - \beta^2 }$ and on the integration contour
\begin{align}
\arg \kappa &= 0 \text{ for $\omega \in \mathbb{R}$  }, \label{arg_k} \\
\label{arg_s}
\arg s &= \left\{
\begin{aligned}
&\pi / 2 \text{ for } 0 < \left| \omega \right| < \omega_{p\perp} \text{ and } \left| \omega \right| > \omega_{p\parallel}, \\
&\pi \text{ for } \omega_{p\perp} < \omega < \omega_{p\parallel}, \\
&0 \text{ for } -\omega_{p\parallel} < \omega < - \omega_{p\perp}.
\end{aligned}
\right.
\end{align}
The branch points, branch cuts and arguments of functions $\kappa \left( \omega \right)$ and $s \left( \omega \right)$ on the whole complex plane are shown in Fig. \ref{transfunctions_complex}. 

\begin{figure*}
    \begin{minipage}{0.49\linewidth}
        \centering
        \includegraphics[scale=1]{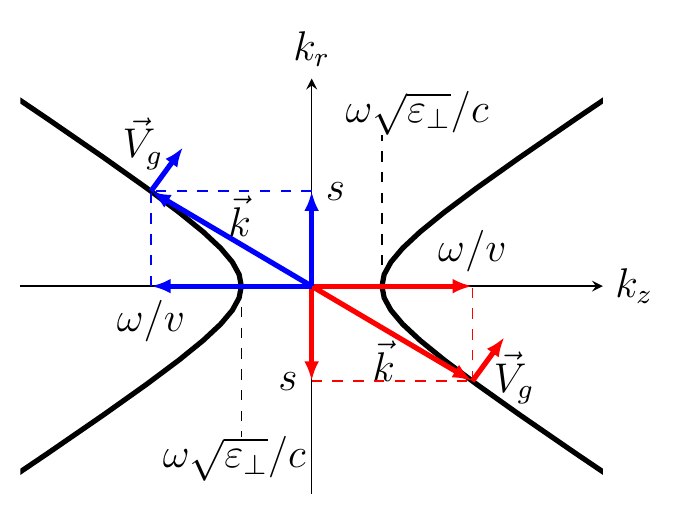}
        \caption{The isofrequency surface of extraordinary waves; $ \vec{V}_g $ (the group velocity vector) is orthogonal to this surface. Area $k_z>0$ (red arrows) corresponds to the case $\omega>0$, area $k_z<0$ (blue arrows) corresponds to $\omega<0$. }
    \label{isofrequency_surface_fig}
    \end{minipage}
    \hfill
    \begin{minipage}{0.49\linewidth}
        \centering
        \includegraphics[width=\linewidth]{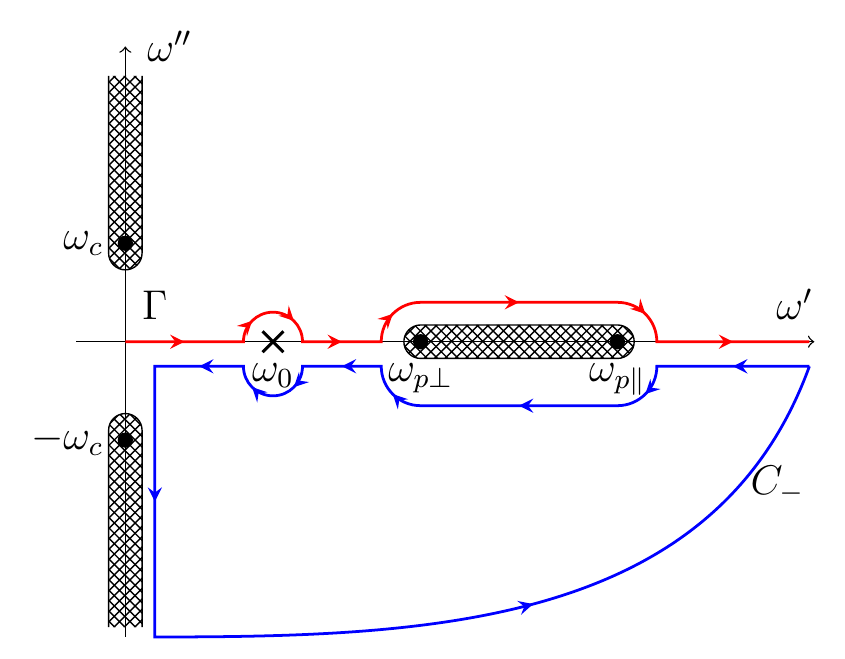}
        \caption{The complex half-plane $\omega$: solid red line means the integration path $\Gamma$; $\omega_0$ is the pole; $\omega_{p\parallel}$, $\omega_{p\perp}$ and $\pm \omega_c$ are the branch points; crosshatched regions are branch cuts. Blue closed line $C_-$ - additional integration path at $\zeta < 0$. } 
    \label{ez_singularity}
    \end{minipage}
\end{figure*}

Note that $s(\omega)$ is imaginary on the real frequencies outside the cuts. Therefore the function $H_0^{\left( 1 \right)} \left( s r \right)$ exponentially decreases with $r$ for such frequencies. These are ``evanescent'' waves that do not take any energy away from the charge.
At the same time, the function
$ s( \omega ) $
is real on the real axis in the ranges
$ \omega_{ p\perp} < |\omega| < \omega_{ p\parallel }$,
which are the radiation frequencies ranges.
Naturally, the integrand~%
\eqref{ez_fourier}
becomes here the quasi-plane extraordinary wave for
$ |s| r \gg 1 $. It is interesting to note that since
$ \mathrm{sgn} \left( s \left(\omega\right) \right) = - \text{sgn} \left( \omega \right) $
in the radiation ranges then the phase velocity $\vec{V}_p$ of the wave is directed to the charge trajectory (i.e. $V_{pr}=\omega s/k^2<0$, $\vec{k}$ is the wave vector). Figure~\ref{isofrequency_surface_fig} shows the isofrequency surface cross-section for the discussed radiated waves. This is a rotation hyperboloid (typical for HMMs~\citep{NaturePhoton, HMM_PQE} and magnetized plasma~\citep{GKT13, GT11}) determined by the equation
\begin{equation}
    \label{isofrequency_surface}
    \frac{k_r^2}{\varepsilon_\parallel} + \frac{k_z^2}{\varepsilon_\perp} - \frac{\omega^2}{c^2} = 0.
\end{equation}

In the problem under consideration $k_z = \omega / v$ and the physically correct solution is $ k_r = s $. Arrows in Fig.~\ref{isofrequency_surface_fig}
illustrate obtaining the physically correct solution for $ s $ (\ref{trans_functions}), (\ref{arg_s}) using the predetermined $ k_z = \omega / v $. As one can see, the group velocity $\vec{V}_g$ (and therefore the Poynting vector which is collinear to  $\vec{V}_g$) is directed away from the charge trajectory, $V_{gr}>0$ (this is physically natural).
%
%

Because of the symmetry property of the integrand $e_{z\omega} \left( -\overline{\omega} \right) = \overline{ e_{z\omega} \left( \omega \right) }$ (the overline means complex conjugation), the expression \eqref{field_integral} can be written as an integral over the semi-axis: 
\begin{equation}
\label{ez_simp}
    E_z = 2\text{Re} \int_0^{+\infty} e_{z\omega} \left(\omega\right) \exp \left( i \zeta \frac{\omega}{v} \right) d\omega.
\end{equation}
Figure~\ref{ez_singularity} shows the location of singularities of the function $e_{z\omega}(\omega)$ (\ref{ez_fourier}) on the complex half-plane $\omega$. The point $\omega_0$ corresponds to the function $1 / F_2 \left(\omega\right)$ pole. One can show that the integration path $\Gamma$ goes around the pole from above and then passes along the upper bank of the cuts, which corresponds to Eq.~\eqref{trans_functions}. 
%

As follows from properties of function $e_{z\omega} \left( \omega \right)$, the integration path $\Gamma$ in expression \eqref{ez_simp} (see Fig.~\ref{ez_singularity}) can be supplemented with a quarter circle of infinite radius in the area $\omega '' >0$ in front of the charge $\left( \zeta >0 \right)$ and infinite quarter circle in the area $\omega '' < 0 $ behind the charge $\left( \zeta < 0 \right)$. The $E_z$ component in front of the charge contains only the the integral $E_z^C$ from $\omega_c$ to $+i\infty$. Behind the charge the supplement of $C_-$ contour leads to the division of the expression \eqref{ez_simp} into the integral $E_z^C$ from $-\omega_c$ to $-i\infty$, integral $E_z^W$ over the branch cut on the real axis and integral $E_z^P$ over an infinitesimal circle around the pole $\omega_0$. Using properties of functions $\kappa \left(\omega\right)$ and $s\left( \omega \right)$, after a series of transformations we obtain
\begin{equation}
\label{ez_qC}
\begin{aligned}
E_z^C &= -\text{sgn} \left( \zeta \right) \frac{ 2 q \sqrt{ 1 - \beta^2 } }{ \pi a v } \int_{\left| \omega_c \right|}^{+\infty} \frac{ e^{ -\omega \left| \zeta \right| / v} }{ | F_2 \left( i \omega \right) |^2 } \\ &\times 
\left\{ 
\begin{aligned}
-\frac{2 |\kappa \left( i \omega \right) | }{ \pi a } J_0 \left( |\kappa \left( i \omega \right) | r \right) \varepsilon_\parallel \left( i \omega \right) \text{ for } r \leq a \\
|s \left( i\omega \right)| \text{Re} \left( H_0^{\left(1\right)} \left( |s \left( i \omega \right) | r \right) F_2 \left( i \omega \right) \right) \text{ for } r>a
\end{aligned}
\right\}d\omega,
\end{aligned} 
\end{equation}
\begin{equation}
\label{ez_wave}
\begin{aligned}
E_z^W &= -\Theta \left(-\zeta\right) \frac{ 4 q \sqrt{ 1 - \beta^2 } }{ a \pi v } \int_{\omega_{p\perp}}^{\omega_{p\parallel}}  \frac{ \cos \left( \omega \zeta / v \right) }{ | F_2 \left( \omega \right) |^2 } \\ &\times 
\left\{
\begin{aligned}
\frac{ 2 \kappa }{ \pi a } | \varepsilon_\parallel | I_0 \left( \kappa r \right) \text{ for } r \leq a \\ 
|s| \text{Im} \left( H_0^{\left(2\right)} \left( |s| r \right) \overline{F_2 \left( \omega \right) } \right) \text{ for } r > a 
\end{aligned}
\right\} d\omega,
\end{aligned}
\end{equation}
\begin{equation}
\label{ez_pole}
\begin{aligned}
E_z^P &= \frac{ \Theta \left( - \zeta \right) 4 q \sqrt{1 - \beta^2} \cos \left( \omega \zeta / v \right) } { C_1 I_1^2\left( \kappa a \right) - C_2 I_0^2\left( \kappa a \right) - C_3 I_0 \left(  \kappa a \right) I_1 \left(  \kappa a \right) } \\ &\times 
\left. \left\{
\begin{aligned} 
\frac{ \kappa }{ a^2 v} I_0 \left( \kappa r \right) \text{ for } r \leq a \\ 
- \frac{ |s| }{ \varepsilon_\parallel a^2 v } \frac{ I_1 \left( \kappa a \right) }{ K_1 \left( |s| a \right) } K_0 \left( |s| r \right) \text{ for } r >a    
\end{aligned}
\right\} \right|_{\omega_0},
\end{aligned} 
\end{equation}
where 
\begin{equation}
\label{coef_p}
\begin{aligned}
&C_1 = \frac{1}{ \omega_{p\parallel}^2 - \omega^2 } \left[ \frac{ \kappa^2 \omega_{p\parallel}^2 }{ \omega } + \frac{ \omega^3 \omega_{p\perp}^2 \left( \omega_{p\parallel}^2 - \omega_{p\perp}^2 \right) }{ v^2 \left( \omega_{p\perp}^2 - \omega^2 \right)^2 } \right] , \\
&C_2 = \frac{ \kappa^2 \omega_{p\perp}^2 \left( \omega_{p\parallel}^2 - \omega^2 \right) }{ \omega \left( 1 - \beta^2 \right) \left( \omega_{p\perp}^2 - \omega^2 \right) \left( \omega^2 + |\omega_c|^2 \right) }, \\
&C_3 = \frac{ 2 \kappa \omega \omega_{p\perp}^2 }{ a \left(1 - \beta^2\right) \left( \omega_{p\perp}^2 - \omega^2 \right) \left( \omega^2 + |\omega_c|^2 \right) }.
\end{aligned}
\end{equation}
The component $E_z^C$ is the quasi-Coulomb field which decreases rapidly with the distance from the charge. Note that the integration in \eqref{ez_qC} starts from point $\omega_c$ (not from $i\delta$ as for the case of charge moving in free space). As it can be shown from the properties of the integrand the interval from $i\delta$ to $\omega_c$ doesn't make a contribution in the quasi-Coulomb field:
\begin{equation}
\label{delta_ez}
\Delta E_z = 2 \text{Re} \int_{i\delta +0}^{\omega_c+0} e_{z\omega} \left( \omega \right) \exp\left({i\zeta \frac{\omega}{v}}\right) d\omega = 0.
\end{equation}

The integration in the second term $E_z^W$  is performed over the frequency range where the function $s \left(\omega\right)$ is real. As was mentioned above, this frequency range is responsible for the wave (or radiation) field. The last component $E_z^P$ exponentially decreases at $|s| r \gg 1$. Correspondingly, the Pointing vector radial component $S_r^P = - c \left(4\pi\right)^{-1} E_z^P H_\varphi^P $ tends to zero exponentially with the distance from the charge trajectory. Thus, the field $E_z^P$ is not a radiation field and can be called a ``plasma train". 

\begin{figure}
\begin{minipage}{0.99\linewidth}
    \begin{minipage}{0.99\linewidth}
        \includegraphics[scale=1]{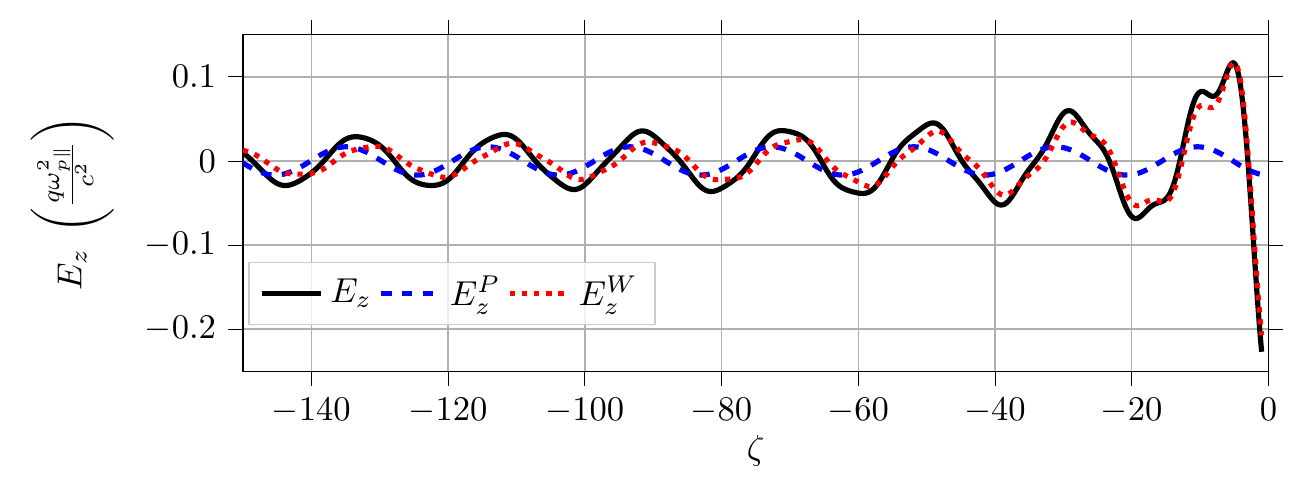}
    \end{minipage}
    \vfill
    \begin{minipage}{0.99\linewidth}
        \includegraphics[scale=1]{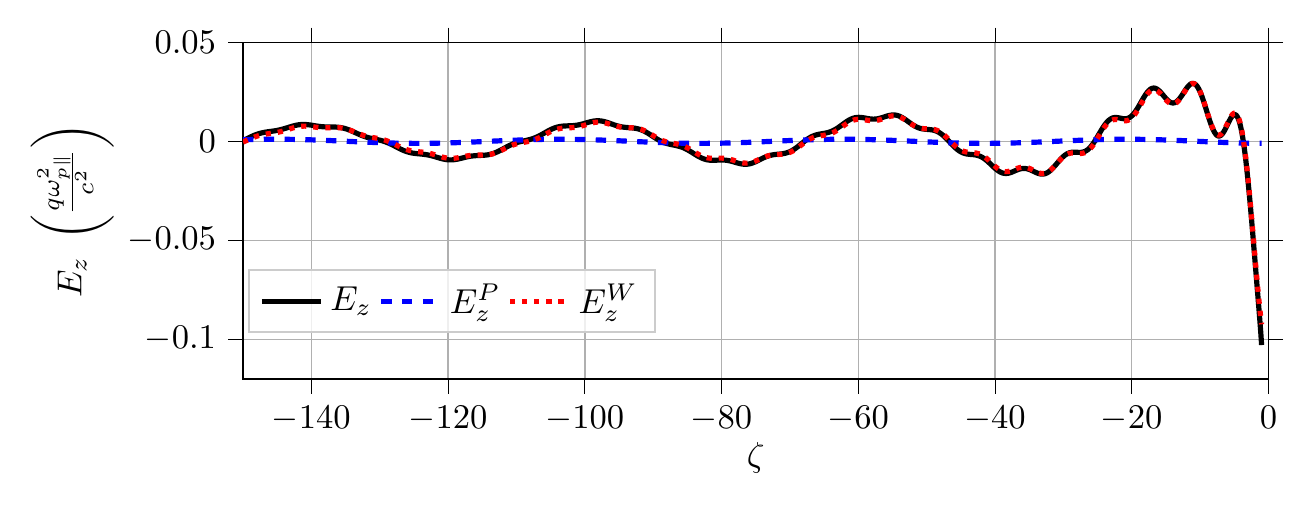}
    \end{minipage}
\end{minipage}
\caption{The full field (black line) $E_z$, wave field (red dotted line) $E_z^W$ and field $E_z^P$ (blue dashed line) dependencies on the $\zeta$ (in units of $c/\omega_{p\parallel}$). The parameters are $a=c/\omega_{p\parallel}$, $r=0.7a$, $\beta=0.99$, $\omega_{p\perp}=0.3\omega_{p\parallel}$ (upper) and $\omega_{p\perp}=0.15\omega_{p\parallel}$ (bottom).}
\label{ez_field}
\end{figure}

The longitudinal component of the total field $\left(E_z\right)$, the radiation field $\left( E_z^W \right)$ and the ``plasma train'' $\left( E_z^P \right)$ are shown in Fig.~\ref{ez_field} for different ``orthogonal'' plasma frequencies. The quasi-Coulomb field decreases rapidly with increasing distance $\zeta$, so the total field behind the charge is mainly determined by the sum of the wave field and ``plasma train''. The ``plasma train'' is sinusoidal because it is excited at a single frequency. The radiation field has a more complex structure because it has a continuous spectrum between $\omega_{p\perp}$ and $\omega_{p\parallel}$. As one can see from Fig.~\ref{ez_field}, the role of the ``plasma train'' reduces with decreasing the ``orthogonal'' plasma frequency. 

\section{Energy loss}
As it is well known, the energy loss per the path length unit is equal to the force acting on the charge with the opposite sign:
\begin{equation}
    \label{energy_loss}
    \frac{dW}{dz_0} = - q \left. E_z \right|_{\zeta=r=0}.
\end{equation}
Due to the Coulomb singularity of the field, we have to transform the integral assuming that $\zeta=0$ and only after that we can find the limit $r \to 0$. It should be underlined that expressions \eqref{ez_qC} - \eqref{ez_pole} can't be used directly for the plane $\zeta=0$ since these formulas are applied to the areas $\zeta >0 $ and $\zeta < 0$. Therefore we have to return to the initial expression \eqref{field_integral}.

Note that the singular summand in the expression (\ref{ez_fourier}) which is proportional to $ \kappa^2 K_0 ( \kappa r) /\omega$ is an odd function of $\omega$ and it does not make a contribution in the integral (\ref{field_integral}) if $\zeta=0$. Therefore we can exclude this singular term. After this, we can put $r \to 0$ and obtain
\begin{equation}
\label{energy_loss_d}
\frac{dW}{dz_0} = \int\nolimits_{-\infty}^{\infty} f\left( \omega \right) d\omega, \; f\left(\omega\right) = - \frac{i q^2}{\pi} \frac{\kappa^2 \left( \omega \right)}{\omega} \frac{F_1\left(\omega\right)}{F_2\left(\omega\right)}.
\end{equation}
Using \eqref{trans_functions}, \eqref{arg_s} one can obtain that $ f\left(\omega\right) = - f\left(-\omega\right) $ on the real axis outside of the cuts, and therefore the ranges $ |\omega| <\omega_{p \perp}$ and  $ |\omega| > \omega_{p \parallel} $ don't contribute to the integral \eqref{energy_loss_d}, with the exception the infinitesimal semicircles near the poles $\pm \omega_0$. Thus, the integral \eqref{energy_loss_d} consists of the contributions of the infinitesimal semicircles around the poles $\left( dW^{\left(p\right)}/dz_0 \right)$ and frequency ranges between branch points $\left( dW^{\left(r\right)} / dz_0 \right)$. Using the residue theorem and the property $f \left( \omega \right) = \overline{ f \left( -\omega \right) }$ on the cuts, finally we obtain the following expression for the energy loss per the path length unit: 
\begin{equation}
\frac{dW}{dz_0} = \frac{dW^{\left(p\right)}}{dz_0} + \frac{dW^{\left(r\right)}}{dz_0},
\label{energy_loss_total}
\end{equation}
\begin{align}
\frac{dW^{\left(p\right)}}{dz_0} =& \frac{ 2 q^2 \kappa^2 \left( \omega_0 \right) }{ a^2 \omega_0 } \left. \frac{1}{  C_2 I_0^2\left( \kappa a \right) - C_1 I_1^2\left( \kappa a \right) +  C_3 I_0 \left(  \kappa a \right) I_1 \left(  \kappa a \right) } \right|_{\omega = \omega_0}, 
\label{energy_loss_pole} \\ 
\frac{dW^{\left(r\right)}}{dz_0} =& \frac{4 q^2 }{ \pi^2 a^2 } \int\nolimits_{\omega_{p\perp}}^{\omega_{p\parallel}}  \frac{ \kappa^2 \left( \omega \right) \left| \varepsilon_\parallel \left(\omega\right) \right| }{ \omega \left| F_2 \left(\omega\right) \right|^2 } d\omega, \label{energy_loss_int_2}
\end{align}
where $\omega_0$ is determined by the equation $F_2 \left( \omega_0 \right) = 0$ and coefficients $C_{1,2,3}$ are \eqref{coef_p}.

The term $dW^{\left(p\right)}/dz_0$ can be named polarization loss, the other term $dW^{\left(r\right)}/dz_0$ corresponds to the radiation loss. Note that another way to obtain the radiation loss is the calculation of the energy flow through an infinitely long cylinder with an axis coinciding with the charge trajectory. Due to the absence of dissipation in the medium, the flow of the radiation energy does not depend on the cylinder radius in contrast to the flow of the ``plasma train'' energy. Because of the pole $\omega_0$ lies in area $\arg \left( s \right) = \pi/2$ the plasma train $E_z^P$ \eqref{ez_pole} decreases exponentially at $r \to \infty$. Thus, the energy flow through the cylinder of infinite radius is equal to the radiation loss. Such a method leads to the same result~\eqref{energy_loss_int_2}. 

\begin{figure}
\includegraphics{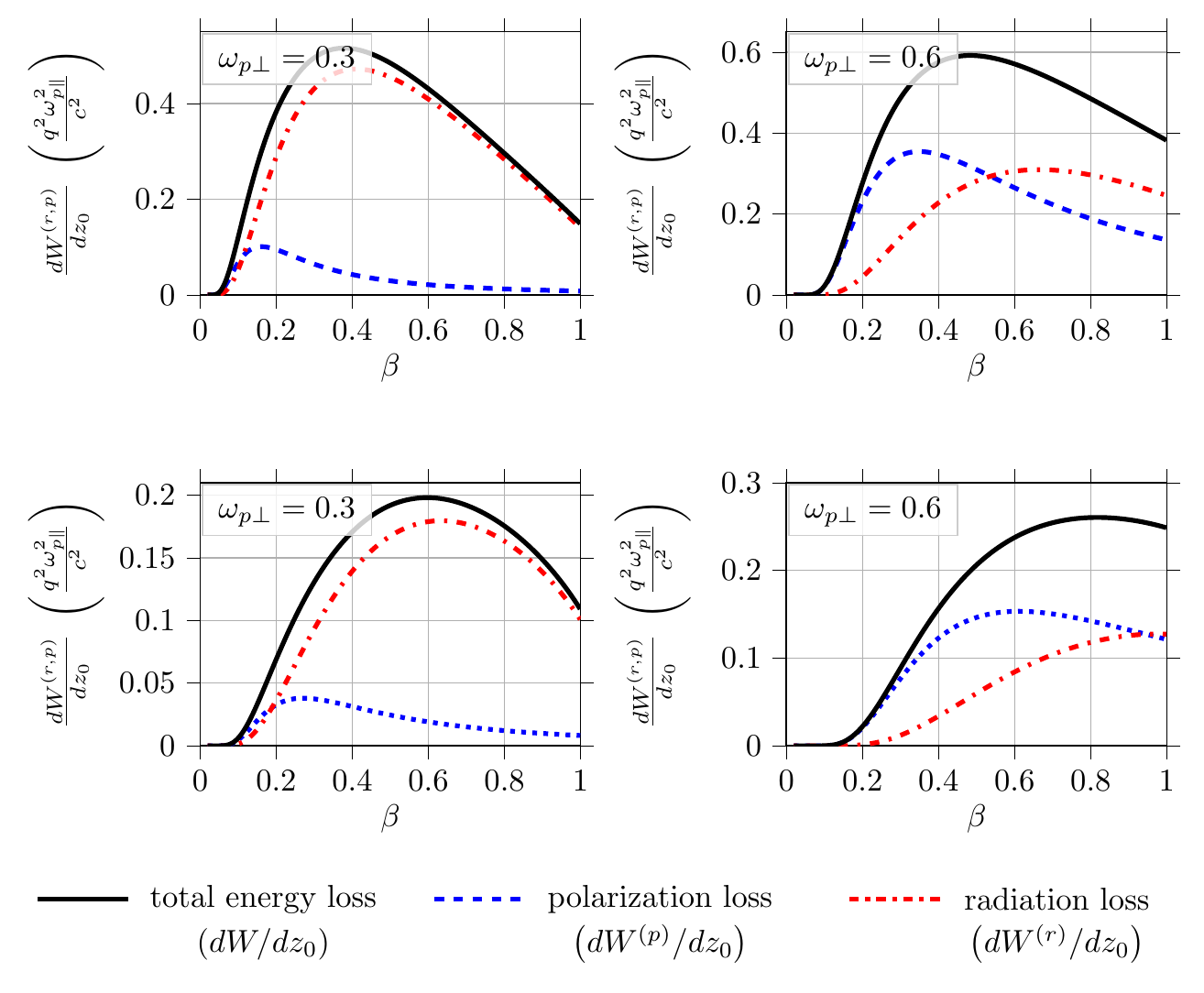}
\caption{Dependence of the energy loss (in units of $q^2 \omega_{p\parallel}^2 / c^2$) on the dimensionless charge velocity $\beta$ at different values of the ``orthogonal'' plasma frequency $\omega_{p\perp}$ (in units of $\omega_{p\parallel}$). Upper line: $a=0.75c/\omega_{p\parallel}$. Bottom line: $a=1.25c/\omega_{p\parallel}$.}
\label{energy_loss_types}
\end{figure}

The total energy loss as well as the radiation loss and the polarization one depends essentially on the charge velocity. These dependencies are shown in Fig.~\ref{energy_loss_types} for different ``orthogonal'' plasma frequencies. As can be seen, the polarization loss can be the main mechanism of the energy loss at sufficiently large values of ``orthogonal'' plasma frequency $\left(\omega_{p\perp}\right)$. It is natural that the vacuum channel narrowing causes an increase in energy loss of both types. However, the larger the channel radius, the lower frequency $\omega_{p\perp}$ is needed for the polarization loss prevalence (in comparison with the energy loss by radiation).

\begin{figure}
    \centering
    \includegraphics{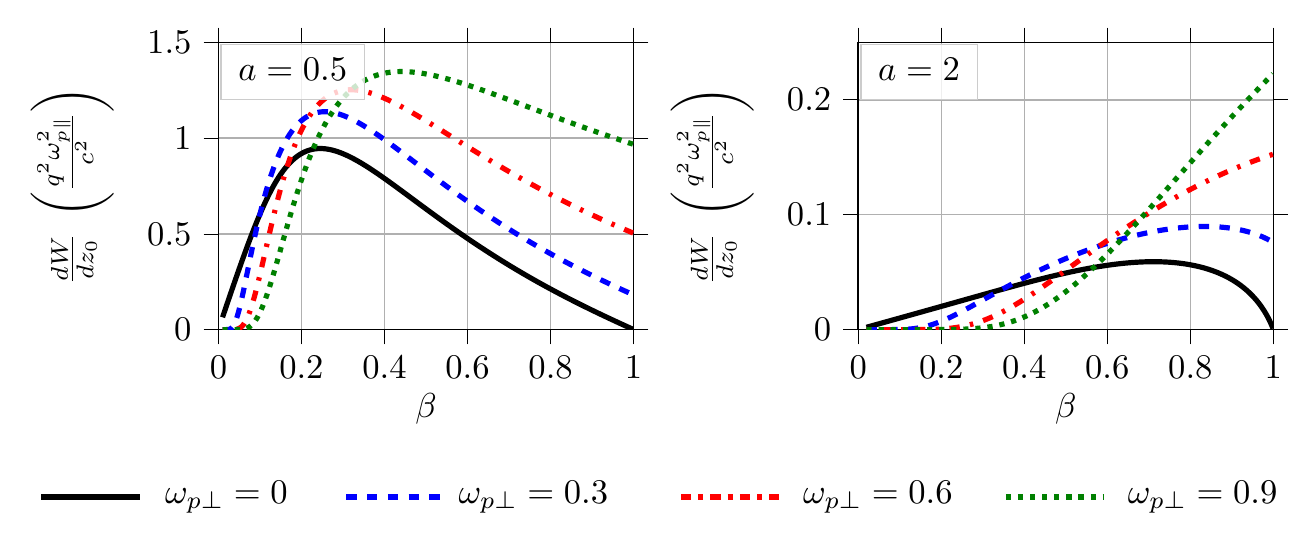}
    \caption{Dependence of the energy loss $ dW/dz_0 $ (in units of $q^2 \omega_{p\parallel}^2 / c^2$) on the dimensionless charge velocity $\beta$ at different values of the channel radius $a$ (in units of $c / \omega_{p\parallel}$). Different lines correspond to different values of ``orthogonal'' plasma frequency $\omega_{p\perp}$ (in units of $\omega_{p\parallel}$).}
\label{energy_loss_channel_radius}
\end{figure}

Figure~\ref{energy_loss_channel_radius} shows the total energy loss $dW/dz_0$ depending on the charge  velocity $\beta$ for different values of $\omega_{p\perp} \big/ \omega_{p \parallel}$ and the channel radius $a$. 
The most important phenomenon is that the energy loss vanishes in the ultrarelativistic limit $\beta \to 1$ at $\omega_{p\perp} \to 0$, i.~e. at $ \varepsilon_\perp \to 1 $. This effect can be explained analytically. In the ultrarelativistic case the polarization loss tends to the following:
\begin{equation}
\label{limit_p}
\lim_{\beta \to 1} \frac{dW^{\left(p\right)} }{dz_0} = \frac{2 q^2}{ \frac{ a^2 \omega_{p\parallel}^2 }{ \omega_{p\perp}^2 - \omega_0^2 } - \frac{ a^4 \omega_0^4 \omega_{p\perp}^2 \left( \omega_{p\parallel}^2 - \omega_{p\perp}^2 \right) }{4 c^2 \left( \omega_{p\parallel}^2 - \omega_0^2 \right) \left( \omega_{p\perp}^2 - \omega_0^2 \right)^2 } }.
\end{equation}
Also, $\omega_0 \to \omega_{p\perp}$ at $\beta \to 1$ and $\omega_{p\perp} \to 0$. So, it can be concluded from the \eqref{limit_p} that $dW^{\left(p\right)}/dz_0$ tends to zero at $\beta \to 1$ under condition $\omega_{p\perp} \to 0$.
The integrand in (\ref{energy_loss_int_2}) tends to the following 
limit at $\beta \to 1$ 
for arbitrary value of $\omega_{p\perp}$:
\begin{multline}
\label{ultrarel}
\lim_{\beta \to 1} \frac{\kappa^2 \left| \varepsilon_\parallel \right| }{ \omega \left| F_2 \right|^2 } = \frac{ \left| \varepsilon_\parallel \right| }{ a^2 \omega } \left[ \left( \frac{ \varepsilon_\parallel }{ a } J_1 \left( \left| s \right| a \right) - \frac{ \left| s \right| }{2} J_0 \left( \left| s \right| a  \right) \right)^2 \right. \\ + \left. \left( \frac{ \varepsilon_\parallel }{ a } Y_1 \left( \left| s \right| a \right) - \frac{ \left| s \right| }{2} Y_0 \left( \left| s \right| a  \right) \right)^2 \right]^{-1},
\end{multline}
where $\left| s \right| = \omega_{p\perp} c^{-1} \left( \omega_{p\parallel}^2 - \omega^2 \right)^{1/2} \left( \omega^2 - \omega_{p\perp}^2 \right)^{-1/2}  $ and $Y_{0,1} \left(x\right)$ are Neumann functions. As one 
can see from (\ref{ultrarel}), the energy loss $dW^{\left(r\right)} / dz_0$ tends to some finite value at $\beta \to 1$ 
if $\omega_{p\perp} \neq 0$. Otherwise, in the case  $\omega_{p\perp} \to 0$, we have  $\left| s \right| \to 0$ and expression (\ref{ultrarel}) tends to zero. Thus, the total energy loss $dW/dz_0 \to 0$ at $\beta \to 1$, $\omega_{p\perp} \to 0$. This result is confirmed by Figure \ref{energy_loss_channel_radius}. 

Although in reality the energy loss is not exactly zero (for various reasons) it however can be dramatically reduced using the media (or artificial materials) having the properties assumed in this paper. This effect can be used for the design of collimators with minimal energy loss.

As was already noted, one of the examples of the medium with the required properties is the hollow electron flux placed in the strong longitudinal magnetic field $ \vec{ H }_0 = H_0 \vec{ e }_z $. The dielectric constant tensor \eqref{e-tensor} with $ \varepsilon_\parallel $ in the form \eqref{epsilons} and $\varepsilon_\perp \approx 1$ takes place if $ \omega_H \gg \omega_{p \parallel}$ where $ \omega_H = e H_0 \big/ ( m c )$ is an electron gyrofrequency and $ \omega_{p \parallel}^2 = 4 \pi e^2 N \big/ m $ is a squared plasma frequency ($e$, $m$ and $N$ are the charge, mass and concentration of electrons, correspondingly).

\section{Conclusion}
We have analyzed the electromagnetic field of the point charge moving through the vacuum channel in the anisotropic non-gyrotropic uniaxial medium with the plasma-like dispersion. It was analytically demonstrated that the field can be devided into the wave (radiation) field, quasi-Coulomb field and ``plasma train''. 

The main attention has been paid to the energy loss investigation. Typical plots of energy loss versus the charge velocity have been demonstrated for the series of values of the vacuum channel radius and the ``orthogonal'' plasma frequency. It has been shown that the relative role of the polarization loss is reduced by decreasing the ``orthogonal'' plasma frequency.

It has been demonstrated and analytically substantiated that the energy loss of the ultrarelativistic charge can be extremely reduced by choosing the appropriate parameters characterizing the medium. In particular, the energy loss tends to zero for the ultrarelativistic charge in the case of $ \varepsilon_{\perp}=1 $. This effect can be of interest for designing collimators in which the bunch energy loss is minimal. The medium with the desired properties can be implemented using certain metamaterials or the hollow electron flux in the strong magnetic field.

\section*{Acknowledgments}

\noindent This work was supported by Russian Science Foundation (Grant No.~18-72-10137).




%

\end{document}